\documentclass[journal]{IEEEtran}
\ifCLASSINFOpdf
\else
   \usepackage{graphicx}
   \graphicspath{{../eps/}}
   \DeclareGraphicsExtensions{.eps}
\fi

\usepackage{amsmath}
\usepackage{amsfonts,amssymb}
\usepackage{amssymb}
\usepackage{wrapfig}
\usepackage{psfrag}
\usepackage{epstopdf}
\usepackage{cite}
\usepackage{graphicx}
\usepackage{subfigure}
\usepackage{threeparttable}
\usepackage{cases}
\usepackage{subeqnarray}
\usepackage{color}
\usepackage{underscore}
\usepackage{verbatim}
\usepackage{bm}
\usepackage{stfloats}

\usepackage{algorithm}
\usepackage{algorithmic}

\begin{document}
\title{Multi-User Communication with Extremely Large-Scale MIMO}

%
\author{Haiquan~Lu
        and
        Yong~Zeng,~\IEEEmembership{Member,~IEEE}\vspace{-0.25em}
\thanks{This work was supported by the National Key R\&D Program of China with Grant number 2019YFB1803400.}
\thanks{H. Lu and Y. Zeng are with the National Mobile Communications Research Laboratory, Southeast University, Nanjing 210096, China. Y. Zeng is also with the Purple Mountain Laboratories, Nanjing 211111, China (e-mail: \{haiquanlu, yong_zeng\}@seu.edu.cn). (\emph{Corresponding author: Yong Zeng.})}
}

\maketitle

\begin{abstract}
Extremely large-scale multiple-input multiple-output (XL-MIMO) communication aims to further boost the antenna size significantly than current massive MIMO systems, for which conventional far-field assumption with uniform plane wave (UPW) model may become invalid. This paper studies the modelling and performance analysis for multi-user XL-MIMO communication. With the spherical wavefront phase modelling, and also by taking into account the variations of signal amplitude and projected aperture across array elements, the performance of the three typical beamforming schemes are analyzed, namely the maximal-ratio combining (MRC), zero-forcing (ZF), and minimum mean-square error (MMSE) beamforming. For the special case of two-users, we analytically show that the signal-to-interference-plus-noise ratio (SINR) of all the three beamforming schemes increases as the channels' correlation coefficient decreases. Furthermore, compared to existing UPW model where inter-user interference (IUI) can only be suppressed in angular domain, XL-MIMO enables a new degree-of-freedom (DoF) for IUI suppression by distance separation, even for users along the same direction. Simulation results are provided to validate the modelling and performance analysis of multi-user XL-MIMO communications.
\end{abstract}

\begin{IEEEkeywords}
Extremely large-scale MIMO, near-/far-field, projected aperture, multi-user communication.
\end{IEEEkeywords}

\IEEEpeerreviewmaketitle
\section{Introduction}
Massive multiple-input multiple-output (MIMO) has become a reality for the fifth-generation (5G) mobile communication networks, where the base station (BS) is typically equipped with at least $64$ antennas~\cite{bjornson2019massive,ngo2013energy}. Looking forward towards beyond 5G (B5G) and 6G, there is momentum in further scaling up the antenna number/size significantly, known as extremely large-scale MIMO (XL-MIMO) \cite{lu2020how,lu2021communicating,gonzalez2021low}, ultra-massive MIMO (UM-MIMO) \cite{akyildiz2016realizing}, or extremely large aperture massive MIMO (xMaMIMO)~\cite{amiri2018extremely}. However, as the antenna size further increases, the commonly used far-field assumption with uniform plane wave (UPW) model may become invalid. Instead, several new channel characteristics need to be considered, such as the near-field radiation with spherical wavefront \cite{zhou2015spherical,gonzalez2021low}, variations of signal amplitude and angle of arrival/departure (AoA/AoD) across array elements \cite{bjornson2020power,torres2020near,hu2018beyond}, and spatial channel non-stationarity \cite{decarvalho2020nonstationarities,ribeiro2021low,marinello2020antenna}.

There are some preliminary efforts on the study of XL-MIMO communications without relying on the conventional UPW models \cite{lu2021communicating,gonzalez2021low,zhou2015spherical,bjornson2020power,torres2020near}. For example, by deviating from the conventional approach that treats the array elements as isotropic and sizeless points, \cite{lu2021communicating} proposed a new approach that takes into account the element's area/aperture, which enables the modelling of the variations of the projected aperture across array elements. We term such model as \emph{projected aperture non-uniform spherical wave} (PNUSW) model. Based on this model, the signal-to-noise ratio (SNR) for single-user communication with the optimal maximal-ratio combining/transmission (MRC/MRT) beamforming was derived in closed-from. In \cite{bjornson2020power}, the power scaling laws and near-field behaviours of single-user XL-MIMO communication was analyzed based on the two-dimensional (2D) channel modelling that only considers the azimuth AoA/AoD, and the extension to two-user system was studied in \cite{torres2020near}. In \cite{marinello2020antenna}, four antenna selection approaches were proposed to improve the energy efficiency for multi-user XL-MIMO communications. Besides, the low-complexity distance-based scheduling and zero-forcing (ZF) precoding schemes were proposed for multi-user XL-MIMO systems in \cite{gonzalez2021low} and \cite{ribeiro2021low}, respectively.

Most of the aforementioned works on XL-MIMO communications treat each array element as isotropic and sizeless points. While practically accurate for small-to-moderate arrays, the isotropic element modelling will lead to results violating physical laws as the array size increases drastically. As an illustration, consider a single-user single-input multiple-output (SIMO) uplink communication system with an isotropic transmitter using power $P$. If the array element at the receiving uniform planar array (UPA) is also modelled isotropically, as the UPA size goes to infinity, its received power will exceed $\frac{P}{2}$ \cite{lu2021communicating}, which is impossible since with isotropic transmitter, the other half of the power will never reach the UPA. Therefore, in this paper, we study the modelling and performance analysis of a multi-user XL-MIMO communication system, where besides the spherical wavefront phase modelling, the variations of signal amplitude and projected aperture across array elements are also explicitly considered. Based on this model, the signal-to-interference-plus-noise ratio (SINR) of the three typical beamforming schemes, i.e., MRC, ZF, and minimum mean-square error (MMSE) beamforming, are analyzed. In particular, for the special case of two users, analytical results are obtained showing that the SINR of all the three beamforming schemes increases as the channels' correlation coefficient decreases. Furthermore, it is found that different from the conventional far-field assumption with the UPW model, the correlation coefficient and hence inter-user interference (IUI) for XL-MIMO communications can be suppressed not only by angle separation, but also by distance separation along the same direction. This thus offers a new degree-of-freedom (DoF) for IUI suppression for multi-user XL-MIMO communications. Simulation results are provided to validate the modelling and
performance analysis of multi-user XL-MIMO communications.

\vspace{-0.3cm}
\section{System Model}\label{sectionSystemModel}
\begin{figure}[!t]
 \setlength{\abovecaptionskip}{-0.1cm}
  \setlength{\belowcaptionskip}{-0.3cm}
  \centering
  \centerline{\includegraphics[width=3.0in,height=2.5in]{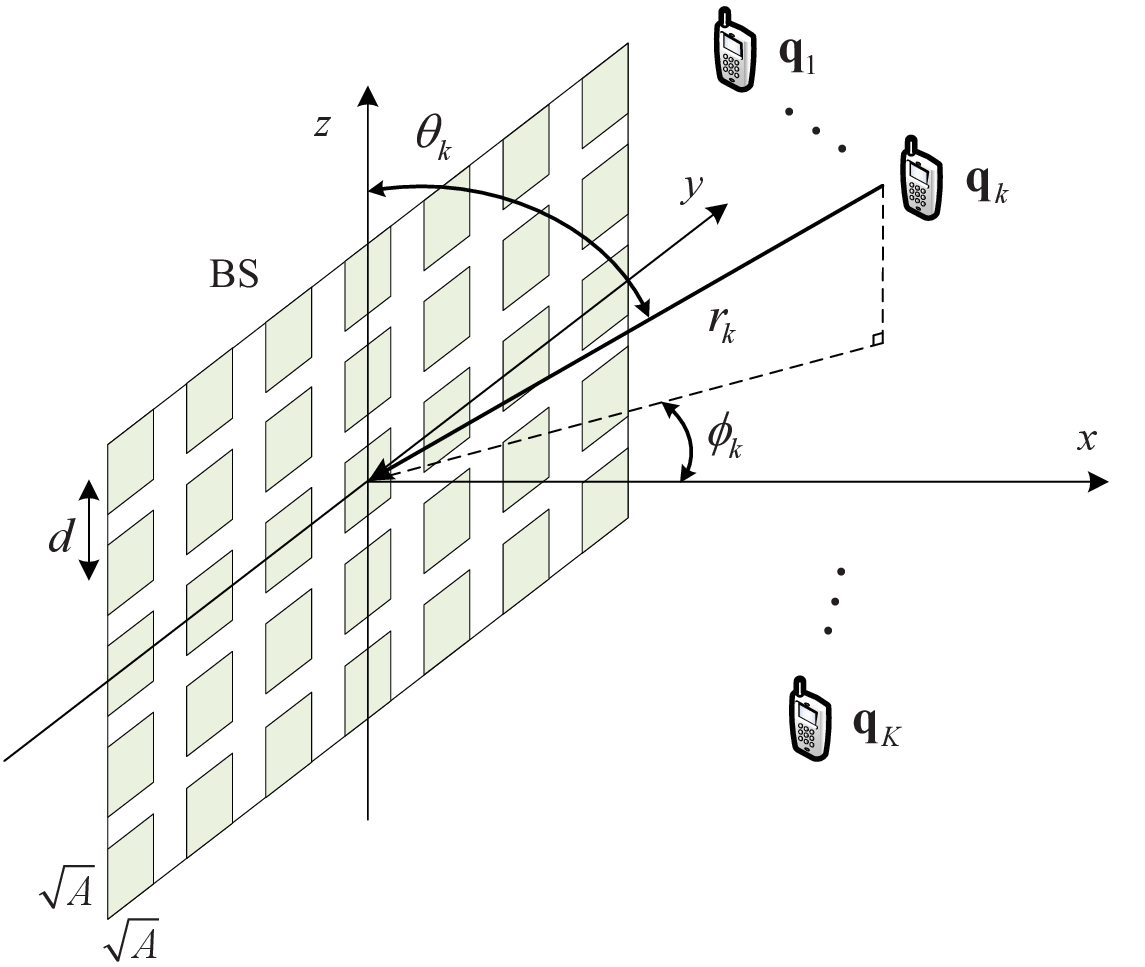}}
  \caption{Multi-user XL-MIMO communication.}
  \label{systemModel}
  \vspace{-0.4cm}
  \end{figure}

As shown in Fig.~\ref{systemModel}, we consider a multi-user communication system, where a BS equipped with an XL-UPA with $M \gg 1$ elements serves $K$ single-antenna users. Without loss of generality, we assume that the UPA is placed on the $y$-$z$ plane and centered at the origin, and $M = {M_y}{M_z}$ with $M_y$ and $M_z$ denoting the number of antenna elements along the $y$- and $z$-axis, respectively. Similar to \cite{lu2021communicating}, instead of modelling each array element as an isotropic and sizeless point, its size is explicitly considered, which is denoted as $\sqrt A \times \sqrt A$. The adjacent elements are separated by $d$, where $d \ge \sqrt A $. Furthermore, we focus on UPA implemented by aperture element such as patch element, which is of low-profile and especially suitable to be mounted on a surface. For convenience, we assume that the aperture efficiency of each element is unity so that its effective aperture is equal to the  physical area $A$.

For notational convenience, we assume that $M_y$ and $M_z$ are odd numbers. The central location of the $\left( {{m_y},{m_z}} \right)$th array element is ${{\bf{w}}_{{m_y},{m_z}}} = {\left[ {0,{m_y}d,{m_z}d} \right]^T}$, where ${m_y} = 0, \pm 1, \cdots , \pm \left( {{M_y} - 1} \right)/2$, ${m_z} = 0, \pm 1, \cdots , \pm \left( {{M_z} - 1} \right)/2$. The physical dimensions of the UPA along the $y$- and $z$-axis are ${L_y} \approx {M_y}d$ and ${L_z} \approx {M_z}d$, respectively. For user $k$, let ${r_k}$ denote its distance with the center of the antenna array, and ${\theta _k} \in \left[ {0,\pi } \right]$ and ${\phi _k} \in \left[ { - \frac{\pi }{2},\frac{\pi }{2}} \right]$ denote the zenith and azimuth angles, respectively. Then the location of user $k$ is ${{\bf{q}}_k} = {\left[ {{r_k}{{\Psi }_k},{r_k}{{\Phi }_k},{r_k}{{\bar \Omega }_k}} \right]^T}$ with ${{\Psi }_k} \buildrel \Delta \over = \sin {\theta _k}\cos {\phi _k}$, ${{\Phi }_k} \buildrel \Delta \over = \sin {\theta _k}\sin {\phi _k}$, and ${{\Omega }_k} \buildrel \Delta \over = \cos {\theta _k}, 1 \le k \le K$. Furthermore, the distance between user $k$ and the center of the $\left( {{m_y},{m_z}} \right)$th antenna element is
\begin{equation}\label{UserAndmthAntennaDistance}
\begin{aligned}
&{r_{k,{m_y},{m_z}}} = \left\| {{{\bf{w}}_{{m_y},{m_z}}} - {{\bf{q}}_k}} \right\|\\
& = {r_k}\sqrt {1 - 2{m_y}{\epsilon _k}{{\Phi }_k} - 2{m_z}{\epsilon _k}{{\Omega }_k} + \left( {m_y^2 + m_z^2} \right)\epsilon _k^2},
\end{aligned}
\end{equation}
where ${\epsilon _k} \buildrel \Delta \over = \frac{d}{{{r_k}}}$. Note that ${r_k} = {r_{k,0,0}}$ and since the array element separation $d$ is on the order of wavelength, in practice, we have $\epsilon_k\ll 1$.

As a theoretical study on the fundamental performance limits, we assume the basic free-space line-of-sight (LoS) propagations. By following the PNUSW model that uses the spherical wavefront phase modelling and also takes into account the variations of signal amplitude and projected aperture across array elements \cite{lu2021communicating}, the channel power gain between user $k$ and the $\left( {{m_y},{m_z}} \right)$th antenna element can be written as
\begin{equation}\label{reducedAntennamChannelPowerGain}
\begin{aligned}
&{g_{{m_y},{m_z}}}\left( {{r_k},{\theta _k},{\phi _k}} \right)\\
& \approx \frac{1}{{4\pi {{\left\| {{{\bf{q}}_k} - {{\bf{w}}_{{m_y},{m_z}}}} \right\|}^2}}}\underbrace {A\frac{{{{\left( {{{\bf{q}}_k} - {{\bf{w}}_{{m_y},{m_z}}}} \right)}^T}{{{\bf{\hat u}}}_x}}}{{\left\| {{{\bf{q}}_k} - {{\bf{w}}_{{m_y},{m_z}}}} \right\|}}}_{{\rm{projected\ aperture}}}\\
& = \frac{{\xi \epsilon _k^2{{\Psi }_k}}}{{{\rm{4}}\pi {{\left[ {1 - 2{m_y}{\epsilon _k}{{\Phi }_k} - 2{m_z}{\epsilon _k}{{\Omega }_k} + \left( {m_y^2 + m_z^2} \right)\epsilon _k^2} \right]}^{\frac{3}{2}}}}},
\end{aligned}
\end{equation}
where ${{\bf{\hat u}}}_x$ denotes the unit vector in the $x$ direction, i.e., the normal vector of each array element, and $\xi  \buildrel \Delta \over = \frac{A}{{{d^2}}} \le 1$ is the \emph{array occupation ratio}, which signifies the fraction of the total UPA plate area that is occupied by the array elements. Therefore, the array response vector for user $k$, denoted as ${{\bf{a}}_k} \in {{\mathbb{C}}^{M \times 1}}$, is formed by the following elements
\begin{equation}\label{newModelArrayResponseVector}
{a_{{m_y},{m_z}}}\left( {{r_k},{\theta _k},{\phi _k}} \right) = \sqrt {{g_{{m_y},{m_z}}}\left( {{r_k},{\theta _k},{\phi _k}} \right)} {e^{ - j\frac{{2\pi }}{\lambda }{r_{k,{m_y},{m_z}}}}},
\end{equation}
where ${m_y} = 0, \pm 1, \cdots , \pm \left( {{M_y} - 1} \right)/2$, ${m_z} = 0, \pm 1,\cdots ,$ $\pm \left( {{M_z} - 1} \right)/2$.

As a comparison, the conventional UPW model is based on the assumption that the array dimension is much smaller than the link distance, such that waves arriving at the receiver array can be well approximated as UPW. Specifically, the $\left( {{m_y},{m_z}} \right)$th element of the array response vector for the UPW model is
 \begin{equation}\label{UPWArrayResponseVector}
 {a_{{m_y},{m_z}}}\left( {{r_k},{\theta _k},{\phi _k}} \right) = \frac{{\sqrt {{\beta _0}} {e^{ - j\frac{{2\pi }}{\lambda }{r_k}}}}}{{{r_k}}}{e^{j\frac{{2\pi }}{\lambda }\left( {{m_y}d{{\Phi }_k} + {m_z}d{{ \Omega }_k}} \right)}},
 \end{equation}
 where ${\beta _0}$ denotes the channel power at the reference distance ${r_0} = 1$~m.

We focus on the uplink communication, while the results can be similarly extended to downlink communication. The received signal at the BS can be expressed as
\begin{equation}\label{BSRaeceivedSignalforUserkWithoutMRC}
{\bf{y}} = {{\bf{a}}_k}\sqrt {{P_k}} {s_k} + \sum\limits_{i = 1,i \ne k}^K {{{\bf{a}}_i}\sqrt {{P_i}} {s_i}}  + {\bf{n}},
\end{equation}
where $P_k$ and $s_k, \forall k$ are the transmit power and information-bearing signal of user $k$, respectively; ${\bf{n}} \sim {\cal C}{\cal N}\left( {0,{\sigma ^2}{{\bf{I}}_M}} \right)$ is the additive white Gaussian noise (AWGN). With linear receive beamforming ${{\bf{v}}_k} \in {{\mathbb{C}}^{M \times 1}}$ applied for user $k$, where $\left\| {{{\bf{v}}_k}} \right\| = 1$, the received SINR is
\begin{equation}\label{ReceivedSINRForkthUser}
{\gamma _k} = \frac{{{{\bar P}_k}{{\left| {{\bf{v}}_k^H{{\bf{a}}_k}} \right|}^2}}}{{\sum\limits_{i = 1,i \ne k}^K {{{\bar P}_i}{{\left| {{\bf{v}}_k^H{{\bf{a}}_i}} \right|}^2}}  + 1}},\ \forall k,
\end{equation}
where ${{\bar P}_k} \buildrel \Delta \over = \frac{{{P_k}}}{{{\sigma ^2}}}$ denotes the transmit SNR for user $k$. The achievable sum rate in bits/second/Hz (bps/Hz) is
\begin{equation}\label{SumRate}
{R_{{\rm{sum}}}} = \sum\limits_{k = 1}^K {{{\log }_2}\left( {1 + {\gamma _k}} \right)}.
\end{equation}

\section{Performance Analysis}\label{sectionPerformanceAnalysis}
In this section, the performance of the three commonly used linear receive beamforming schemes is analyzed, i.e., MRC, ZF, and MMSE beamforming.

1) \emph{MRC beamforming}: For user $k$, the MRC receive beamforming vector is
 \begin{equation}\label{MRCReceiveBeamformingVector}
 {{\bf{v}}_{{\rm{MRC}},k}}= \frac{{{\bf{a}}_k}}{{\left\| {{\bf{a}}_k} \right\|}},\ \forall k.
 \end{equation}
 The SINR in \eqref{ReceivedSINRForkthUser} can then be expressed as
 \begin{equation}\label{MRCReceivedSINRForkthUser}
 \begin{aligned}
 {\gamma _{{\rm{MRC}},k}} = \frac{{{{\bar P}_k}{{\left\| {{{\bf{a}}_k}} \right\|}^2}}}{{\sum\limits_{i = 1,i \ne k}^K {{{\bar P}_i}\frac{{{{\left| {{\bf{a}}_k^H{{\bf{a}}_i}} \right|}^2}}}{{{{\left\| {{{\bf{a}}_k}} \right\|}^2}}}}  + 1}} = {{\bar P}_k}{\left\| {{{\bf{a}}_k}} \right\|^2}\left( {1 - {\alpha _{{\rm{MRC}},k}}} \right),
 \end{aligned}
 \end{equation}
 where ${{\bar P}_k}{\left\| {{{\bf{a}}_k}} \right\|^2}$ is the single-user SNR as if there were no IUI, ${\alpha _{{\rm{MRC,}}k}} = \frac{{\sum\limits_{i = 1,i \ne k}^K {{{\bar P}_i}{\rho _{ki}}} {{\left\| {{{\bf{a}}_i}} \right\|}^2}}}{{\sum\limits_{i = 1,i \ne k}^K {{{\bar P}_i}{\rho _{ki}}} {{\left\| {{{\bf{a}}_i}} \right\|}^2} + 1}}$ with $0 \le {\alpha _{{\rm{MRC}},k}} \le 1$ can be interpreted as the SNR loss factor due to the IUI, and ${\rho _{ki}} \buildrel \Delta \over = \frac{{{{\left| {{\bf{a}}_k^H{{\bf{a}}_i}} \right|}^2}}}{{{{\left\| {{{\bf{a}}_k}} \right\|}^2}{{\left\| {{{\bf{a}}_i}} \right\|}^2}}}$ with $0\leq {\rho _{ki}}\leq 1$ denotes the channels' correlation coefficient between user $k$ and $i$.

2) \emph{ZF beamforming}: The ZF receiver is designed such that the IUI is completely removed, which requires $M \ge K$. For user $k$, we have
\begin{equation}\label{ZFReceiverBeamformingVector}
{\bf{v}}_{{\rm ZF},k}^H{{{\bf{\bar A}}}_k} = {{\bf{0}}_{1 \times \left( {K - 1} \right)}},
\end{equation}
 where ${{{\bf{\bar A}}}_k} = \left[ {{{\bf{a}}_1}, \cdots ,{{\bf{a}}_{k - 1}},} \right.\left. {{{\bf{a}}_{k + 1}}, \cdots ,{{\bf{a}}_K}} \right]$. Therefore, the ZF receive beamforming vector for user $k$ is
 \begin{equation}\label{ZFReceiveBeamformingVectorNormalized}
 {{\bf{v}}_{{\rm{ZF}},k}} = \frac{{\left( {{{\bf{I}}_M} - {{{\bf{\bar A}}}_k}{{\left( {{\bf{\bar A}}_k^H{{{\bf{\bar A}}}_k}} \right)}^{ - 1}}{\bf{\bar A}}_k^H} \right){{\bf{a}}_k}}}{{\left\| {\left( {{{\bf{I}}_M} - {{{\bf{\bar A}}}_k}{{\left( {{\bf{\bar A}}_k^H{{{\bf{\bar A}}}_k}} \right)}^{ - 1}}{\bf{\bar A}}_k^H} \right){{\bf{a}}_k}} \right\|}},\ \forall k,
 \end{equation}
 where ${{{\bf{I}}_M} - {{{\bf{\bar A}}}_k}{{\left( {{\bf{\bar A}}_k^H{{{\bf{\bar A}}}_k}} \right)}^{ - 1}}{\bf{\bar A}}_k^H}$ is the projection matrix into the space orthogonal to the columns of ${{{{\bf{\bar A}}}_k}}$ \cite{brown2012practical}. By substituting \eqref{ZFReceiveBeamformingVectorNormalized} into \eqref{ReceivedSINRForkthUser}, we have
 \begin{equation}\label{ZFReceivedSINRForkthUser}
 \begin{aligned}
 {\gamma _{{\rm{ZF}},k}} &= {{\bar P}_k}{\bf{a}}_k^H\left( {{{\bf{I}}_M} - {{{\bf{\bar A}}}_k}{{\left( {{\bf{\bar A}}_k^H{{{\bf{\bar A}}}_k}} \right)}^{ - 1}}{\bf{\bar A}}_k^H} \right){{\bf{a}}_k}\\
  &= {{\bar P}_k}{\left\| {{{\bf{a}}_k}} \right\|^2}\left( {1 - {\alpha _{{\rm{ZF,}}k}}} \right),
 \end{aligned}
 \end{equation}
 where ${\alpha _{{\rm{ZF,}}k}} = \frac{{{\bf{a}}_k^H{{{\bf{\bar A}}}_k}{{\left( {{\bf{\bar A}}_k^H{{{\bf{\bar A}}}_k}} \right)}^{ - 1}}{\bf{\bar A}}_k^H{{\bf{a}}_k}}}{{{{\left\| {{{\bf{a}}_k}} \right\|}^2}}}$ with $0 \le {\alpha _{{\rm{ZF}},k}} \le 1$. Similarly, ${\alpha _{{\rm{ZF,}}k}}$ accounts for the SNR loss factor caused by the cancellation of IUI with ZF beamforming.

3) \emph{MMSE beamforming}: The SINR in \eqref{ReceivedSINRForkthUser} is a generalized Rayleigh quotient with respect to ${\bf v}_k$, which is maximized by the linear MMSE beamforming, i.e.,
\begin{equation}\label{MMSEReceiveBeamformingVectorNormalized}
{{\bf{v}}_{{\rm{MMSE}},k}} = \frac{{{\bf{C}}_k^{ - 1}{{\bf{a}}_k}}}{{\left\| {{\bf{C}}_k^{ - 1}{{\bf{a}}_k}} \right\|}},\ \forall k,
\end{equation}
where ${{\bf{C}}_k} = \sum\limits_{i \ne k}^K {{{\bar P}_i}} {{\bf{a}}_i}{\bf{a}}_i^H + {{\bf{I}}_M}$ is the interference-plus-noise covariance matrix. The resulting SINR for user $k$ is
\begin{equation}\label{MMSEReceivedSINRForkthUser}
{\gamma _{{\rm{MMSE}},k}} = {{\bar P}_k}{\bf{a}}_k^H{\bf{C}}_k^{ - 1}{{\bf{a}}_k} = {{\bar P}_k}{\left\| {{{\bf{a}}_k}} \right\|^2}\left( {1 - {\alpha _{{\rm{MMSE}},k}}} \right),
\end{equation}
where ${\alpha _{{\rm{MMSE}},k}} = \frac{{{\bf{a}}_k^H\left( {{{\bf{I}}_M} - {\bf{C}}_k^{ - 1}} \right){{\bf{a}}_k}}}{{{{\left\| {{{\bf{a}}_k}} \right\|}^2}}}$ can be interpreted as the SNR loss factor caused by the IUI with MMSE beamforming. It can be shown that ${{{\bf{I}}_M} - {\bf{C}}_k^{ - 1}}$ is a positive semidefinite matrix, and thus we have $0 \le {\alpha _{{\rm{MMSE}},k}} \le 1$.

To gain further insights, we consider the special case of two users, i.e., $K=2$, for which the SINR with MRC, ZF and MMSE beamforming for user 1 respectively reduce to
\begin{equation}\label{twoUsersMRCSINRUser1}
\hspace{-1ex}{\gamma _{{\rm{MRC}},1}} = \frac{{{{\bar P}_1}{{\left\| {{{\bf{a}}_1}} \right\|}^2}}}{{{{\bar P}_2}{{\left\| {{{\bf{a}}_2}} \right\|}^2}{\rho _{12}} + 1}} = {{\bar P}_1}{\left\| {{{\bf{a}}_1}} \right\|^2}\left( {1 - \frac{{{{\bar P}_2}{{\left\| {{{\bf{a}}_2}} \right\|}^2}{\rho _{12}}}}{{{{\bar P}_2}{{\left\| {{{\bf{a}}_2}} \right\|}^2}{\rho _{12}} + 1}}} \right).
 \end{equation}
\begin{equation}\label{twoUsersZFSINRUser1}
 {\gamma _{{\rm{ZF}},1}} = {{\bar P}_1}{\left\| {{{\bf{a}}_1}} \right\|^2}\left( {1 - {\rho _{12}}} \right).
 \end{equation}
\begin{equation}\label{twoUsersMMSESINRUser1}
{\gamma _{{\rm{MMSE}},1}} = {{\bar P}_1}{\left\| {{{\bf{a}}_1}} \right\|^2}\left( {1 - \frac{{{{\bar P}_2}{{\left\| {{{\bf{a}}_2}} \right\|}^2}{\rho _{12}}}}{{1 + {{\bar P}_2}{{\left\| {{{\bf{a}}_2}} \right\|}^2}}}} \right).
\end{equation}
It is observed from \eqref{twoUsersMRCSINRUser1}-\eqref{twoUsersMMSESINRUser1} that with any given channel powers ${\left\| {{{\bf{a}}_1}} \right\|^2}$ and ${\left\| {{{\bf{a}}_2}} \right\|^2}$, the SINR of all the three beamforming schemes increases as the channels' correlation coefficient $\rho _{12}$ decreases. Based on \eqref{twoUsersMRCSINRUser1}-\eqref{twoUsersMMSESINRUser1}, it can also analytically show the widely known conclusions that ${\gamma _{{\rm{MMSE}},1}} \ge {\gamma _{{\rm{ZF}},1}}$ and ${\gamma _{{\rm{MMSE}},1}} \ge {\gamma _{{\rm{MRC}},1}}$. Furthermore, by comparing \eqref{twoUsersMRCSINRUser1} and \eqref{twoUsersZFSINRUser1}, when ${\rho _{12}} \le 1 - \frac{1}{{{{\bar P}_2}{{\left\| {{{\bf{a}}_2}} \right\|}^2}}}$, we have ${\gamma _{{\rm{ZF}},1}} \ge {\gamma _{{\rm{MRC}},1}}$. By substituting \eqref{newModelArrayResponseVector} into the correlation coefficient, we have \eqref{correlationCoefficient} shown at the top of the next page. While the closed-form expression of ${{{\left\| {{{\bf{a}}_k}} \right\|}^2}}$ was obtained in our preliminary work \cite{lu2021communicating}, that for the numerator of \eqref{correlationCoefficient} is in general difficult to obtain. However, for the conventional UPW model, it follows from \eqref{UPWArrayResponseVector} that the correlation coefficient reduces to
\newcounter{mytempeqncnt1}
\begin{figure*}
\normalsize
\setcounter{mytempeqncnt1}{\value{equation}}
\begin{align}  \label{correlationCoefficient}
{\rho _{ki}} = \frac{{{{\left| {\sum\limits_{{m_z} =  - \frac{{{M_z} - 1}}{2}}^{ \frac{{{M_z} - 1}}{2}} {\sum\limits_{{m_y} =  - \frac{{{M_y} - 1}}{2}}^{ \frac{{{M_y} - 1}}{2}} {\sqrt {{g_{{m_y},{m_z}}}\left( {{r_k},{\theta _k},{\phi _k}} \right){g_{{m_y},{m_z}}}\left( {{r_i},{\theta _i},{\phi _i}} \right)} {e^{j\frac{{2\pi }}{\lambda }\left( {{r_{k,{m_y},{m_z}}} - {r_{i,{m_y},{m_z}}}} \right)}}} } } \right|}^2}}}{{{{\left\| {{{\bf{a}}_k}} \right\|}^2}{{\left\| {{{\bf{a}}_i}} \right\|}^2}}}.
\end{align}
\hrulefill
\vspace{-0.5cm}
\end{figure*}
\begin{equation}\label{correlationCoefficientUPW}
\begin{aligned}
{\rho _{ki}} &= \frac{1}{{{M^2}}}{\left| {\sum\limits_{{m_z} =  - \frac{{{M_z} - 1}}{2}}^{\frac{{{M_z} - 1}}{2}} {\sum\limits_{{m_y} =  - \frac{{{M_y} - 1}}{2}}^{\frac{{{M_y} - 1}}{2}} {{e^{ - j2\pi \left( {{m_y}\bar d\Delta {\Phi _{ki}} + {m_z}\bar d\Delta {\Omega _{ki}}} \right)}}} } } \right|^2}\\
 &= \frac{1}{{{M^2}}}{\left| {\frac{{\sin \left( {\pi {M_y}\bar d\Delta {\Phi _{ki}}} \right)}}{{\sin \left( {\pi \bar d\Delta {\Phi _{ki}}} \right)}}} \right|^2}{\left| {\frac{{\sin \left( {\pi {M_z}\bar d\Delta {\Omega _{ki}}} \right)}}{{\sin \left( {\pi \bar d\Delta {\Omega _{ki}}} \right)}}} \right|^2},
\end{aligned}
\end{equation}
where $\Delta {\Phi _{ki}} \buildrel \Delta \over = {\Phi _k} - {\Phi _i}$, $\Delta {\Omega _{ki}} \buildrel \Delta \over = {\Omega _k} - {\Omega _i}$, and $\bar d \buildrel \Delta \over = \frac{d}{\lambda }$. Note that for users located along the same direction, i.e., $\Delta {\Phi _{ki}} = \Delta {\Omega _{ki}} = 0$, we have ${\rho _{ki}} = 1$ for UPW model, while it is generally smaller than one for the PNUSW model, as will be shown in the following section.

\section{Simulation Results}\label{sectionSimulationResults}
In this section, simulation results are provided to compare the PNUSW and conventional UPW models for multi-user XL-MIMO communications. Unless otherwise stated, the reference SNR of all users is ${{\bar P}_k}{\beta _{\rm{0}}}= 50$~dB, $\forall k$, and the antenna separation is set as $d = \frac{\lambda }{2}=0.0628$~m. The size of each array element is $A = \frac{{{\lambda ^2}}}{{4\pi }}$.

\begin{figure}[!h]
  \centering
  \centerline{\includegraphics[width=3.0in,height=1.8in]{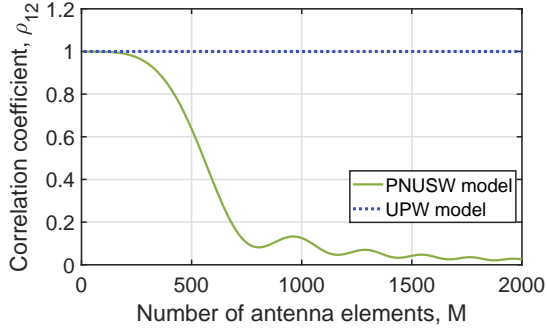}}
  \caption{Correlation coefficient ${\rho _{12}}$ versus antenna number for the PNUSW and UPW models. The users are located along the same direction.}
  \label{correlationCoefficientVersusAntennaNumberProjected}
  \vspace{-0.5cm}
 \end{figure}

Fig.~\ref{correlationCoefficientVersusAntennaNumberProjected} compares the correlation coefficient ${\rho _{12}}$ versus antenna number $M$ for the PNUSW and UPW models. The two users are located along the same direction but with different distances to the center of the antenna array, which are given by $\left( {{r_1},{\theta _1},{\phi _1}} \right) = \left( {25{\rm{~m}},\frac{\pi }{2},0} \right)$ and $\left( {{r_2},{\theta _2},{\phi _2}} \right) = \left( {250{\rm{~m}},\frac{\pi }{2},0} \right)$, respectively. The number of antennas along the $y$-axis is fixed as $M_y = 10$, and $M$ is varied by varying $M_z$. It is observed that the correlation coefficient with the UPW model is always equal to one, which is expected since $\Delta {\Phi _{12}} = \Delta {\Omega _{12}} = 0$ in \eqref{correlationCoefficientUPW}. By contrast, the correlation coefficient with the PNUSW model generally decreases as $M$ increases. This implies that the increase of antenna number helps to reduce the correlation and hence IUI for multi-user XL-MIMO communications.

 \begin{figure}[!h]
  \centering
  \centerline{\includegraphics[width=3.0in,height=1.8in]{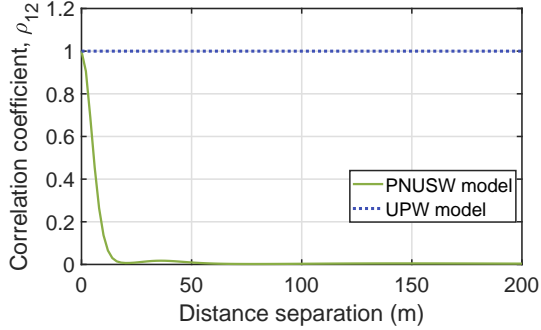}}
  \caption{Correlation coefficient ${\rho _{12}}$ versus distance separation for the PNUSW and UPW models. The users are located along the same direction.}
  \label{correlationCoefficientVersusTheDistanceSeparationprojected}
  \vspace{-0.3cm}
 \end{figure}
As a further illustration, Fig.~\ref{correlationCoefficientVersusTheDistanceSeparationprojected} compares the correlation coefficient ${\rho _{12}}$ versus distance separation $\left| {{r_1} - {r_2}} \right|$ for the PNUSW and UPW models. The location of user 1 is $\left( {{r_1},{\theta _1},{\phi _1}} \right) = \left( {50{\rm{~m}},\frac{\pi }{2},0} \right)$, and the direction of user 2 is $\left( {\theta_2 ,\phi_2 } \right) = \left( {\frac{\pi }{2},0} \right)$. The number of antenna elements is $M_y=M_z=200$. It is observed that as the distance separation increases, the correlation coefficient with the PNUSW model decreases in general, while that with the UPW model is always equal to one. This implies that the IUI can be suppressed not only by direction separation as in the conventional UPW model, but also by distance separation even for users along the same direction, which offers a new DoF for IUI suppression with XL-MIMO.

\begin{figure}[!t]
\centering
\subfigure[PNUSW model]{
\begin{minipage}[t]{0.5\textwidth}
\centering
\centerline{\includegraphics[width=3.0in,height=1.8in]{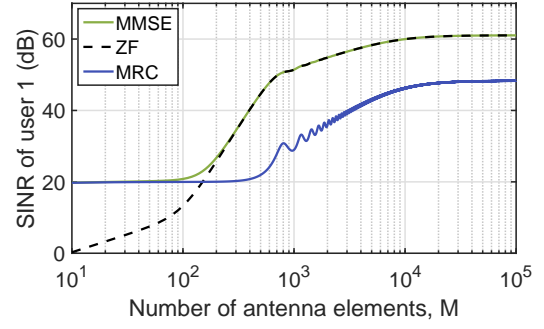}}
\end{minipage}
}
\subfigure[UPW model]{
\begin{minipage}[t]{0.5\textwidth}
\centering
\centerline{\includegraphics[width=3.0in,height=1.8in]{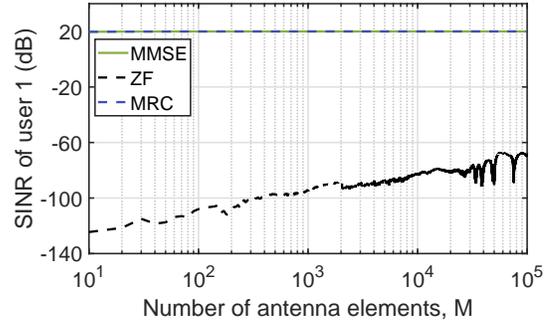}}
\end{minipage}
}
\caption{SINR of user 1 with MRC, ZF and MMSE beamforming versus antenna number for the PNUSW and UPW models. The users are located along the same direction.}
\label{twoUsersSNRMRCZFMMSEVersusAntennaNumberSameAngle}
\vspace{-0.5cm}
\end{figure}

Fig.~\ref{twoUsersSNRMRCZFMMSEVersusAntennaNumberSameAngle} shows the SINR of user 1 with MRC, ZF and MMSE beamforming versus antenna number for the PNUSW and UPW models. The number of antenna elements along the $y$-axis is fixed as $M_y = 10$. The location of two users are $\left( {{r_1},{\theta _1},{\phi _1}} \right) = \left( {25{\rm{~m}},\frac{\pi }{2},0} \right)$ and $\left( {{r_2},{\theta _2},{\phi _2}} \right) = \left( {250{\rm{~m}},\frac{\pi }{2},0} \right)$. It is firstly observed that for the PNUSW model, the SINR of user 1 with MRC and MMSE beamforming is approximately a constant for moderate $M$. This is because that the correlation coefficient is relatively large in this regime, as can be seen from Fig.~\ref{correlationCoefficientVersusAntennaNumberProjected}, for which user 1 suffers from the severe interference from user 2. In addition, the relatively large correlation coefficient also results in the poor performance for ZF beamforming, as can be inferred from \eqref{twoUsersZFSINRUser1}. As $M$ further increases, benefiting from XL-MIMO in reducing the correlation coefficient, the SINR of user 1 with MRC, ZF and MMSE beamforming increases and eventually approaches to a constant value, as expected. By contrast, for UPW model, the performance of MRC and MMSE beamforming remains constant since the correlation coefficient in \eqref{twoUsersMRCSINRUser1} and \eqref{twoUsersMMSESINRUser1} is always equal to one as both users are along the same direction. Furthermore, the ZF beamforming with UPW model
results in extremely poor SINR since it is unable to separate the users that are located along the same direction.

\begin{figure}
\centering
\subfigure[PNUSW model]{
\begin{minipage}[t]{0.5\textwidth}
\centering
\centerline{\includegraphics[width=3.0in,height=1.8in]{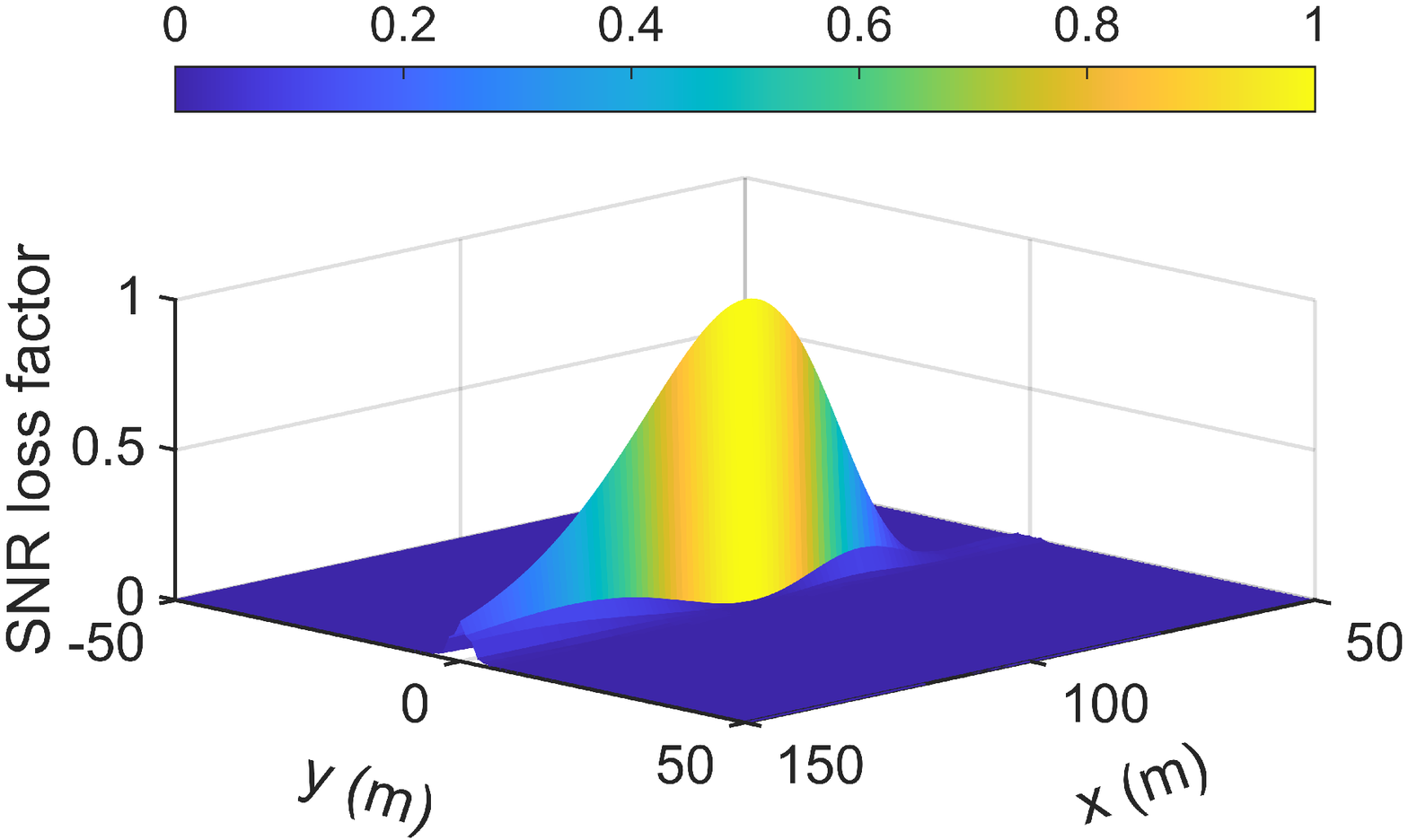}}
\end{minipage}
}
\subfigure[UPW model]{
\begin{minipage}[t]{0.5\textwidth}
\centering
\centerline{\includegraphics[width=3.0in,height=1.8in]{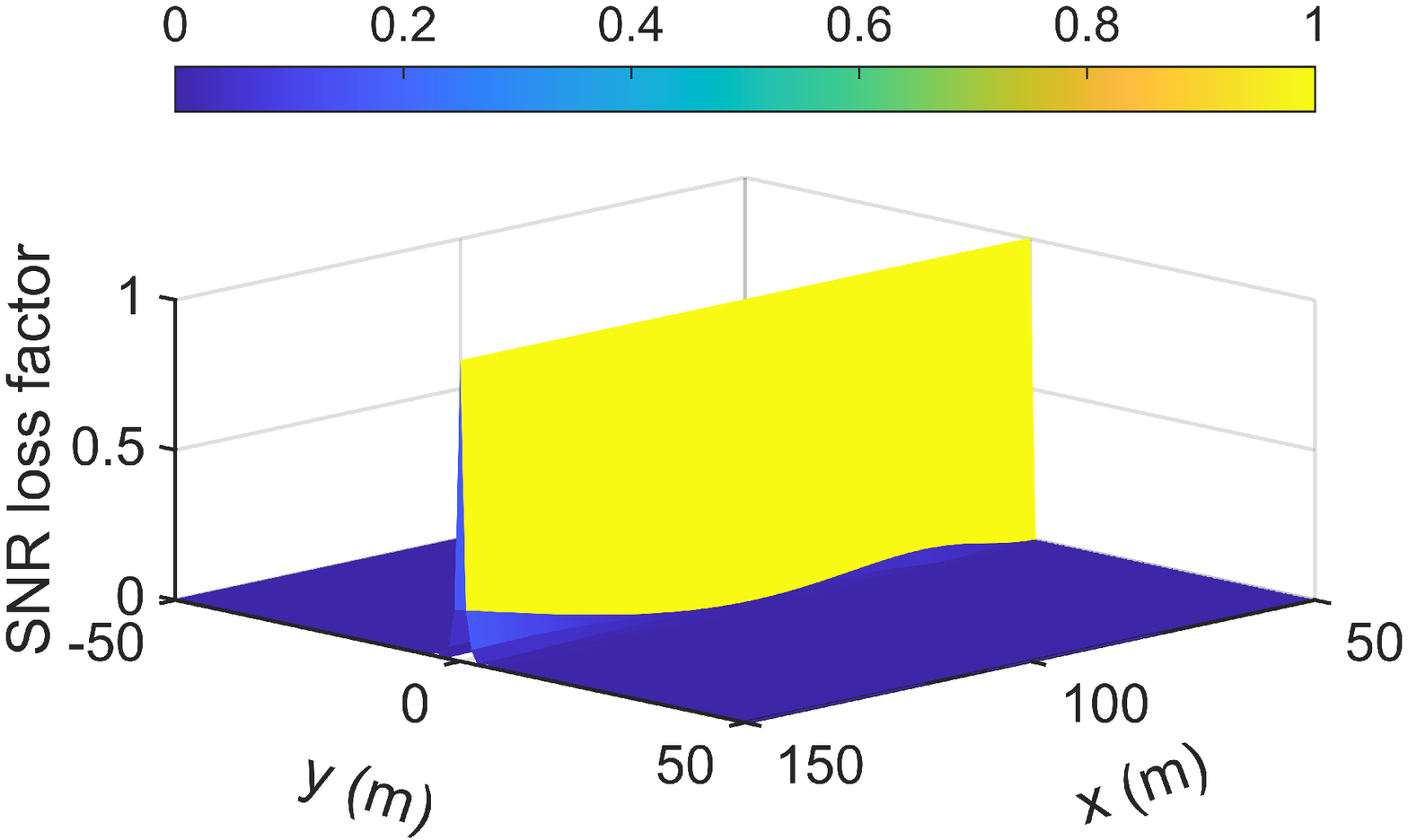}}
\end{minipage}
}
\caption{SNR loss factor of user 1 with MMSE beamforming versus the location of user 2 for the PNUSW and UPW models.}
\label{NewUPWMMSEtwoUsersPerformanceLossversusUserLocation}
\vspace{-0.3cm}
\end{figure}

By fixing the location of user 1 to $\left( {{r_1},{\theta _1},{\phi _1}} \right) = \left( {100{\rm{~m}},\frac{\pi }{2},0} \right)$, Fig.~\ref{NewUPWMMSEtwoUsersPerformanceLossversusUserLocation}
shows its SNR loss factor with MMSE beamforming (i.e., ${\alpha _{\rm{MMSE,}1}}$) by varying the location of user 2 on the $x$-$y$ plane. The number of antenna elements is $M_y = M_z = 200$. It is observed that with the UPW model, the SNR loss is severe for all locations of user 2 that have the same direction as user 1 (i.e., along the $x$-axis). By contrast, with the PNUSW model, notable SNR loss is observed only for user 2 located sufficiently close to user 1, thanks to the DoF offered by XL-MIMO for IUI suppression by the distance separation along the same direction.

 \begin{figure}[!t]
  \centering
  \centerline{\includegraphics[width=3.0in,height=2.1in]{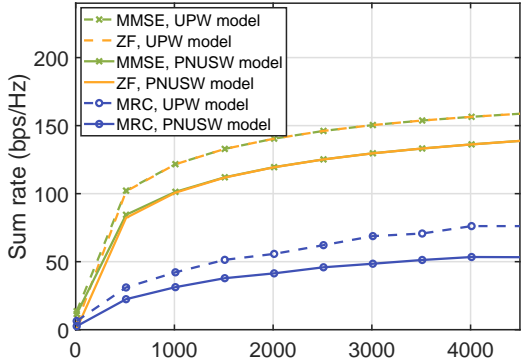}}
  \caption{Sum rate versus antenna number with MMSE, ZF and MRC beamforming for the PNUSW and UPW models. $K=10$ users are randomly distributed in the region given by ${r_k} \in \left[ {50,100} \right]$~m, ${\theta _k} \in \left[ {0,\frac{\pi }{3}} \right]$, ${\phi _k} \in \left[ {\frac{\pi }{6},\frac{\pi }{3}} \right]$, $\forall k$.}
  \label{sumRateVersusNumberofAntennaElementsMMSEZFMRC}
  \vspace{-0.5cm}
 \end{figure}

Last, Fig.~\ref{sumRateVersusNumberofAntennaElementsMMSEZFMRC} compares the sum rate with MMSE, ZF and MRC beamforming versus antenna number $M$ for the PNUSW and UPW models. We assume that $K=10$ users are randomly distributed in the region given by ${r_k} \in \left[ {50,100} \right]$~m, ${\theta _k} \in \left[ {0,\frac{\pi }{3}} \right]$, ${\phi _k} \in \left[ {\frac{\pi }{6},\frac{\pi }{3}} \right]$, $\forall k$. It is firstly observed that UPW model in general over-estimates the sum rate predicted by the PNUSW model, and the gap becomes more significant as $M$ increases. This can be explained by the following two reasons. First, due to the ignorance of variations of projected aperture across array elements, UPW model in fact exaggerates the effective aperture to intercept the impinging wave, especially for waves with inclined incidence. Second, with the UPW model, all array elements are assumed to have one identical AoA for each user, which thus exaggerates the angle separation for different users while each of them in fact has different AoAs with respect to different regions of the XL-MIMO. As a result, the IUI is under-estimated with the conventional UPW model, and hence the sum rate is over-estimated.

\vspace{-0.3cm}
\section{Conclusion}\label{sectionConclusion}
This paper studied the modelling and performance analysis of multi-user communication with XL-MIMO. With the spherical wavefront phase modelling and by explicitly considering the variations of signal amplitude and projected aperture across array elements, the SINR of the typical MRC, ZF and MMSE beamforming schemes were analyzed. It was found that there exists a new DoF for IUI suppression by distance separation along the same direction for multi-user XL-MIMO communications. Simulation results were presented to validate the modelling and performance analysis of multi-user XL-MIMO communications.


\bibliographystyle{IEEEtran}
\bibliography{refXLMIMOMultiUser}
\end{document}